\documentclass[pra]{revtex4}
\newtheorem{allBlock}{Conjecture}
\newtheorem{no9U}[allBlock]{Theorem}
\newtheorem{no3}[allBlock]{Theorem}
\usepackage{graphicx}
\usepackage{amssymb}

\begin{document}
\title{Non-optimality of unitary operations for dense coding}
\author{Michael R. Beran$^1$, and Scott M. Cohen$^{1,2}$}
\email{cohensm@duq.edu}
\affiliation{$^1$Department of Physics, Duquesne University, Pittsburgh,
Pennsylvania 15282\\$^2$Department of Physics, Carnegie-Mellon University,
Pittsburgh, Pennsylvania 15213
}

\begin{abstract}
One of the primary goals of information theory is to provide limits on the amount of information it is possible to send through various types of communication channels, and to understand the encoding methods that will allow one to achieve such limits. An early surprise in the study of \textit{quantum} information theory was the discovery of dense coding, which demonstrated that it is possible to achieve higher rates for communicating classical information by transmitting quantum systems, rather than classical ones. To achieve the highest possible rate, the transmitted quantum system must initially be maximally entangled with another that is held by the receiver, and the sender can achieve this rate by encoding her messages with unitary operations. The situation where these two systems are not maximally entangled has been intensively studied in recent years, and to date it has appeared as though unitary encoding might well be optimal in all cases. Indeed, this optimality of unitary operations for quantum communication protocols has been found to hold under far more general conditions, extending well beyond the special case of dense coding. Nonetheless, we here present strong numerical evidence supported by analytical arguments that indicate there exist circumstances under which one can encode strictly more classical information using dense coding with non-unitary, as opposed to unitary, operations.
\end{abstract}

\maketitle
\section{Introduction}\label{sec:intro}
It is of fundamental importance in quantum information theory to understand the transmission of information from one party to another when that information is carried by quantum systems; that is, by systems whose behavior manifests quantum characteristics, such as the peculiarly quantum correlations known as entanglement \cite{Schrodinger1,EPR,Bell}. One of the most important aspects of this goal is the study of quantum channel capacities, specifically, the amount of information that can be conveyed by various types of channels. These channels may operate under specific restrictions, and may be supplemented by additional resources that are available to the parties. For example, the channel will generally restrict the size of the system (dimension of its Hilbert space); it may be noisy due to insufficient isolation of the system from the environment during transmission; the parties may only be able to use the channel once, or alternatively, an unlimited number of times. In addition, they may be able to exchange classical messages, either in both directions or in only one direction and not the other; they may have prior shared entanglement; etc. 

We are interested here in the question of finding the best encoding method for the transmission of classical information through quantum channels under various constraints. Specifically, we wish to know when a protocol that is restricted to unitary operations for encoding individual messages can nonetheless be optimal. To be clearer, suppose Alice performs a quantum operation ${\cal A}_j$ on system $\cal Q$ (which we assume is initally in a pure state) to encode message $j$, and then after transmission through the channel, Bob measures $\cal Q$ to decode her message. Can Alice achieve the highest possible transmission rate of classical information by using unitary operations (which leave $\cal Q$ in a pure state) for her encoding?

If there is no prior shared entanglement between Alice and Bob, it is readily shown that this question has an affirmative answer no matter what other restrictions are included in their protocol. The reason is that if she encoded messages as mixed states, this will never be better than if she instead encoded individual messages as the (pure) eigenstates of those mixed states \cite{SchuWest}, since the latter contains additional ``which state" information missing in the mixed state messages. This conclusion will then hold for all channels, both classical and quantum, no matter how many uses, $\eta$, of that channel the parties are allowed to use, including the limit ${\eta}\rightarrow\infty$, corresponding to what is commonly known as the (classical) capacity of the channel. 

We are then left with the question of using channels assisted by entanglement. Does the above conclusion still hold? In this case, $\cal Q$ consists of two entangled subsystems, $A$ held by Alice and $B$ held by Bob. To encode her message Alice is only able to operate on $A$, having no access to $B$, but Bob will decode by measuring the entire system $\cal Q$. The reason this scenario differs so significantly from the case where Alice can operate on the entire system $\cal Q$ (or equivalently, if $A$ and $B$ are initially in a product state) is that she can now utilize --- and manipulate --- the correlations between the subsystems to increase the amount of information she is able to encode. If there are no correlations to begin with, a restriction to unitary operations will in no way limit the range of possibilities. When entanglement is present, on the other hand, this is no longer the case. To illustrate this point, we note that Alice's unitary operations do not change the amount of entanglement. In contrast, by using non-unitary operations she can, for example, change a maximally entangled state to one which is completely unentangled, and can even change a state with very little entanglement to one that is maximally entangled (though not deterministically). When there is prior shared entanglement, then, it is not at all clear that pure state encoding can always do at least as well as encoding with mixed states: \textit{Alice now has the ability to encode with mixed states whose eigenstates cannot be reached by unitary encoding}.

If $\cal Q$ starts out in a pure entangled state, of which the parties have many copies, and if they can use a noisy channel an arbitrarily large number of times, it has been shown in \cite{BennettEntPRL,BennettEnt} that unitary encoding is still optimal. The case where the parties use a perfect (noiseless) channel, on the other hand, is commonly known as dense coding \cite{BennettDense}. Several papers \cite{TerhalDense,Piani,Winter,Bowen,Hiroshima,Bruss} considered the condition that arbitrarily many copies of a (possibly mixed) entangled state are available and arbitrarily many uses of the noiseless channel are allowed, some \cite{TerhalDense,Piani,Winter} showing that Alice can send the maximum amount of information by using unitaries to encode her messages, while others \cite{Bowen,Hiroshima,Bruss} assumed this restriction to unitary encoding from the outset%(it should be noted that in the former case, there is an optimal protocol that begins with a more general operation, the same for all messages, which optimizes the shared state and is followed by unitary encoding of the specific chosen message)
. Indeed, to our knowledge with only a single exception, every study of dense coding to date has either assumed unitary encoding (examples include \cite{BennettDense,Bowen,Hiroshima,Bruss,Mozes,ChineseDense,Agrawal,Pati,Guo,BosePlenio,GerjuoyDense}) or found that such encoding is optimal  \cite{TerhalDense,Piani,Winter} (the one exception is \cite{ourDense}, in which we raised the question of whether unitaries are optimal for dense coding, but did not address this issue further).

In this paper, we argue that in spite of this preponderance of evidence, there exist circumstances under which unitary encoding is strictly less than optimal. We consider the case when Alice and Bob share a single copy of a system in a pure, partially entangled state, Alice is allowed a single use of a perfect quantum channel to send her part of this system to Bob, and the operations used by Alice to encode her chosen message must be such that Bob is always able to determine with certainty the message she has sent. Such protocols are known as deterministic dense coding. We present convincing numerical evidence that unitary encoding is \textit{not} optimal for a range of initial states of the shared, entangled resource. We then give analytical arguments that strongly support this conclusion.

\section{Deterministic dense coding}
Deterministic dense coding with non-maximally entangled states has been studied previously in \cite{Mozes,ourDense,Ji}. The initial setup is as follows: Alice and Bob know the state of their shared entangled system, described by Hilbert space ${\cal H}_{AB}={\cal H}_A\otimes{\cal H}_B$ of dimension $d\times d$ (both subsystems have dimension $d$). This state may be written in its Schmidt decomposition \cite{NielsenChuang}, 
\begin{equation}\label{Psi0}
	\vert\Psi_0\rangle = \sum_{m=0}^{d-1} \sqrt{\lambda_m}\vert m\rangle_A\vert m\rangle_B,
\end{equation}
with $\lambda_0\ge \lambda_1\ge \cdots \ge\lambda_{d-1}$ real and positive, and $\sum_m\lambda_m=1$. Alice and Bob together choose a set of states to represent the messages she will later send. As explained below, any given message will be represented by one or more pure states. In order for Bob to always be able to determine with certainty which message Alice has sent him, the states representing any given message must be orthogonal to those representing any other one.

To implement the protocol, Alice begins by choosing a message and then encodes it by performing the corresponding local operation on her part of their shared system. She then sends system $A$ to Bob, after which he measures on the combined system ${\cal Q}=A\otimes B$ to determine Alice's message. No other communication between them is allowed.

After choosing message $j$, Alice performs operation ${\cal A}_j$ on her system $A$. If ${\cal A}_j$ is unitary, the effect will be a transformation of the initial state to $(U_j\otimes I_B)|\Psi_0\rangle$, where $I_{B(A)}$ is the identity operator on ${\cal H}_{B(A)}$. Otherwise, we will imagine that Alice brings in an ancillary system $a_j$ (of dimension $\kappa_j$) in some fixed initial state, performs unitary $U_j$ on the combination of systems $a_j$ and $A$, after which she can either measure the ancilla or throw it away. Let us suppose she measures it and obtains outcome $l$ (throwing it away does not change things in any important way). The effect on system $A$ will in general be a non-unitary operation, represented by the Kraus operator $K_{jl}$ \cite{Kraus}. As she will want to make her own choice of which message to send, rather than to allow the random outcome of her measurement on $a_j$ to determine the message, the two of them must agree that all these operations $K_{jk}$ (with fixed $j$ and $k=1, \cdots, \kappa_j$) will collectively represent message $j$. Note that unitarity of $U_j$ implies that for each $j$,
\begin{equation}\label{complete}
	I_A = \sum_{k=1}^{\kappa_j} K_{jk}^\dagger K_{jk}.
\end{equation}
In addition, if ${\cal A}_j$ is unitary, then $\kappa_j=1$.

In order for Bob to determine with certainty which message has been sent, the states representing message $j$,
\begin{equation}
	\vert\Psi_{jk}\rangle = (K_{jk}\otimes I_B)\vert\Psi_0\rangle,
\end{equation}
must be orthogonal to all states corresponding to other messages, $\vert\Psi_{j^\prime k^\prime}\rangle$ for $j^\prime \ne j$, a requirement that may be succinctly written as
\begin{equation}\label{lamOrtho}
	\textrm{Tr}(K_{jk}\Lambda K_{j^\prime k^\prime}^\dagger)=0~~~~~(j\ne j^\prime),
\end{equation}
We will refer to this condition as ``$\Lambda$-orthogonality" between the two Kraus operators, with $\Lambda$ a diagonal matrix having diagonal elements $\lambda_0,\lambda_1,\cdots,\lambda_{d-1}$, in that order. Note that each of Alice's messages is here represented by a mixed state, a probabilistic mixture of the pure states $|\Psi_{jk}\rangle$ with fixed $j$.

We will assume without loss of generality that for fixed $j$, states $\vert\Psi_{jk}\rangle$ are linearly independent, so that $\kappa_j$ determines the dimension of the subspace in ${\cal H}_{AB}$ spanned by the states $\vert\Psi_{jk}\rangle$ (for fixed $j$; $\kappa_j$ is then referred to as the Kraus rank of ${\cal A}_j$). The requirement of orthogonality for $j^\prime\ne j$ restricts the choice of $\{{\cal A}_j\}$, and depending on the Schmidt coefficients, $\lambda_m$, determines the maximum number of messages the parties can include in their ``codebook" \cite{Mozes,ourDense}. 

A general condition that restricts Alice's set of operations when they are linearly independent is that
\begin{equation}\label{kappa}
	\sum_j\kappa_j \le d^2.
\end{equation}
This follows directly from pair-wise orthogonality of states corresponding to any two different messages, since there can be no more than $d^2$ linearly independent states, and points to an interesting trade-off between unitary and non-unitary operations. Allowing Alice to use non-unitaries means that she has a (far) wider range of operations to choose from, and one might therefore suspect she can do better by using this larger toolbox. However, since any unitary message is represented by a single state, whereas each non-unitary is represented by two or more, then with all else being equal, it will be easier to fit more unitary messages into the $d^2$-dimensional Hilbert space. As a result of these two opposing trends, it is rather difficult to predict in any particular situation whether or not it will be beneficial to allow non-unitary operations. It is shown in Appendix \ref{app:d2} that there can be no such benefit when $d=2$, the case of two qubits. In Section \ref{sec:d4}, we present numerical results for $d=3$, which indicate there is again no benefit to using non-unitary operations. However, we then present compelling numerical evidence that this is no longer the case in $d=4$: it appears quite clear that unitary encoding is not optimal in at least one region of the parameter space of Schmidt coefficients. Following this, we give analytical arguments that strongly support this conclusion.

\section{Dense coding with higher-dimensional systems}\label{sec:d4}
While it is possible to rigorously prove that non-unitary operations are of no benefit for the case of $d=2$ (see Appendix \ref{app:d2}), a generalization of this proof to higher dimensions, even just to $d=3$, appeared to present a significant challenge. We therefore turned our attention to numerical studies, which we describe in the following sections. 

\subsection{Numerical methods}
As our numerical approach is precisely the same as was used in \cite{Mozes} to study the case of unitary encoding -- we have simply modified their code, which they graciously shared with us, to allow for inclusion of non-unitary messages -- the reader is referred to their paper for a description of these methods. We note in particular the discussion in their Section VI-A regarding the reliability of their numerical results, in reference to the fact that this procedure is a search method, requiring an initial starting point to be supplied. One could worry, as they did, that unless one supplied a ``good enough" initial guess, one's numerical search might be unable to find a solution that nonetheless exists. This is especially important since it would require a prohibitively large amount of computing time to thoroughly search the relevant, and rather large, parameter space representing allowable sets of ${\cal N}$ operators (${\cal N}$ is equal to the number of messages, $N$, when restricting to unitary encoding). However, they observed that for any choice of $|\Psi_0\rangle$ and ${\cal N}$, only two cases occur: either (1) the procedure converges to a solution for every starting point one uses, or (2) it never finds a solution no matter how many different starting points one tries. They used these observations to argue that it is very likely their numerical search always finds a solution whenever one exists. We have observed precisely the same pattern with non-unitaries and therefore concur with the authors of \cite{Mozes} that this approach does indeed appear to be a very reliable method of capturing the true properties of these systems.

Our primary goal is to see if it is ever possible to send a larger number of messages with non-unitary, as compared to unitary, encoding. Since each non-unitary message consists of two or more Kraus operators, we have that ${\cal N}>N$ for this case, implying that the parameter space one is searching through to find a given number of messages is necessarily larger than when restricting to unitaries. One might well expect, then, that it is \textit{at least as difficult computationally} to find $N$ non-unitary messages as it is to find the same number of unitary ones. That is, if for a given $|\Psi_0\rangle$, $N$ non-unitaries are found but $N$ unitaries cannot be, it seems unlikely that an increase of computational resources would allow one to then find a set of $N$ unitary messages. Therefore, if a region is found numerically where non-unitary encoding does better than unitary, we can have a degree of confidence in these results.

%[This latter point about a larger parameter space is given added significance for the following reason: Note first that the need to fit all messages into the $d^2$-dimensional Hilbert space, Eq.~(\ref{kappa}), leaves a great deal of freedom to choose the Kraus ranks of sets of $N$ messages. For example with $N=7$ and $d=3$, we could have (1) $7$ unitaries, or (2) $6$ unitaries and one message of Kraus rank $2$, or (3) $6$ unitaries and one message of Kraus rank $3$, or (4) $5$ unitaries and two messages each of Kraus rank $2$. A priori, one does not know which, if any, of these choices is best. Our approach was to choose for each message large enough Kraus ranks to include all possibilities in one run of the search algorithm. In the above example, then, we searched for $5$ unitaries, a $6^{\textrm{\small{th}}}$ message having $2$ Kraus operators, and a $7^{\textrm{\small{th}}}$ message having $3$ Kraus operators. Note, however, that by Eq.~(\ref{kappa}), the $5$ Kraus operators (representing messages $6$ and $7$) cannot be linearly independent since they also must be orthogonal to the $5$ unitaries in this $d^2=9$ dimensional space. While allowing for all these searches to be done in a single calculation, this approach increases the size of the parameter space one is searching through, presumably making it even more difficult computationally to find solutions. Nonetheless, it is still the case that we either always find solutions, or we never do.]

Below, we present our results, which will include a discussion of phase boundaries in the space represented by the $d-1$ independent Schmidt coefficients, $\lambda_0, \cdots, \lambda_{d-2}$. The phases which these boundaries delineate are regions of this $\lambda$-space in which the maximum number of messages is some fixed number, $N$. We will refer, for example, to the (unitary) $6$-boundary as the surface which separates the region in which $6$ (unitary) messages can be found, from that region in which we can only find $5$. We note that our numerically generated data points for each boundary are found to lie on smooth surfaces in all cases.

\subsection{Numerical results}\label{sec:numrslts}
We began our numerical studies by looking at the case $d=3$, which with a restriction to unitary operations, was completely mapped in Fig.~$1$ of Mozes, et.al. \cite{Mozes}. After first reproducing this phase diagram for unitary encoding, we then mapped the non-unitary case, finding precisely the same boundaries for both cases. Although not a rigorous proof, this is a strong indication that non-unitary encoding does not offer any advantage over encoding that is unitary in $d=3$. While this could have been viewed as a somewhat negative result, we felt it was nonetheless of interest, since we had not previously known what to expect. There was now some reason to anticipate the same sort of result in higher dimensions. Nonetheless, our study of the $d=4$ case showed a very different, and we believe extremely interesting, behavior. 

We have completely mapped the phase diagram for deterministic dense coding in $4\times4$ using only unitary operations. The results of this mapping are quite interesting in their own right and will be reported elsewhere \cite{BeranU}. We then looked at non-unitary encoding, and found at least one region in which $9$ messages are found where the numerical search fails to find $9$ unitary messages. There are two additional regions, one for $7$ and another for $14$ messages, which appear to exhibit this same behavior. These latter two regions are relatively small, however, making it more difficult to argue away the possibility they are numerical artifacts. We feel quite strongly, on the other hand, that the $N=9$ region represents a real effect.

\begin{figure}[h]
\includegraphics{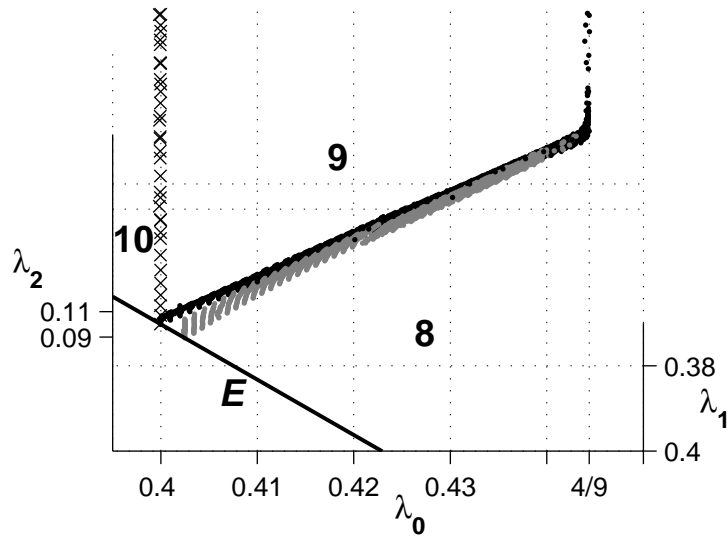}
\caption{\label{fig:9K}Phase boundaries for dense coding with a $4$-by-$4$ entangled state. The vertical line of black data points (crosses) represents the boundary between regions of $10$ (to the left) and $9$ (to the right) unitary messages; the vertical line of black dots further to the right, the boundary between $9$ and $8$ unitaries, with this set of points continuing downward and to the left separating the region of $9$ unitary messages from that of $9$ non-unitary messages; and below this, the gray points separating the region of $9$ non-unitary messages from that of $8$ unitaries. The solid black line extending down and to the right in the bottom portion of the figure represents the edge (labeled in the figure, and referred to in the text, as $E$) defined by the two conditions, $\lambda_1=\lambda_0$ and $\lambda_3=\lambda_2$. Note that two or these boundaries, one separating $9$ and $10$ unitaries and the other separating $9$ unitaries from $9$ non-unitaries, meet at the point along this edge where $\lambda_0=0.4$ and $\lambda_2=0.1$.}
\end{figure}

In Fig.~\ref{fig:9K}, we have labeled various regions corresponding to different numbers of messages; specifically, we see a region where $10$ unitary messages can be found, one with $9$ unitaries, and another with $8$ unitaries. In addition, there is a small but significant area, between the lower sets of black and gray boundary points, where the numerical calculations are unable to find $9$ unitary messages, but do (easily) find $9$ messages when Alice is allowed to use non-unitary operations.

The boundary between the region allowing $10$ unitaries and that allowing only $9$ unitaries is found to be a flat plane (vertical line of black crosses in the figure) located at $\lambda_0=0.4=d/N$ ($N=10$). It has been rigorously shown elsewhere that even allowing non-unitary operations, this value of $\lambda_0$ is an upper bound; that is, it is only possible to send $N$ messages deterministically using dense coding with a $d\times d$ system if $\lambda_0\le d/N$ \cite{ourDense}.

The boundary to the right of the region of $9$ unitaries is quite unusual, exhibiting a transition where it appears to abruptly depart from a flat plane at the upper-bound $\lambda_0=d/N=4/9$ to what appears to be another (almost) flat plane angled down and to the left, which precisely meets the unitary $10$ boundary along the edge $E$ (solid black line extending down and to the right in the bottom portion of the figure) where $\lambda_0=0.4$. While several other boundaries display curvature both in $d=3$ and $d=4$, this is the only boundary we have seen that displays such a sharp transition. Close inspection, however, does seem to show that the `break' is not a discontinuity in slope, but is slightly gradual in nature.

Note the clear separation between the boundary below the region of $9$ unitaries and that above the region of $8$ unitaries. These two boundaries, which smoothly approach each other and meet near (or possibly right at) the break in the former, individually meet $E$ at $\lambda_0=0.4$ and $0.4025$, respectively. While this may seem to be a small separation, it is nonetheless quite significant, being $50$ times greater than the resolution of our numerical calculations.

\subsection{Analytical supporting arguments}
In an effort to more deeply understand these results, we focused our attention on the edge $E$ where $\lambda_1=\lambda_0$ and $\lambda_3=\lambda_2$, which is rather special in the following way: Note that it is always possible to think of a $d=4$ system as a combination of two qubits, and the entangled $4\times 4$ system Bob and Alice are using as $4$ qubits, two with Alice and two with Bob. Along $E$ it is readily seen that one of Alice's qubits is maximally entangled with one of Bob's (more specifically, this pair is in the Bell-$0$ state, defined by $\sqrt{2}|B_0\rangle=|0\rangle_{A_1}|0\rangle_{B_1}+|1\rangle_{A_1}|1\rangle_{B_1}$), while the other pair of qubits is partially entangled.
From the numerical results, we see that $E$ is special in another way -- the point where the unitary $9$ boundary meets the $10$ boundary also lies on $E$, specifically at $\lambda_1=\lambda_0=0.4$ and $\lambda_3=\lambda_2=0.1$, which we may represent as
\begin{equation}
	\Lambda=\Lambda_*=\left(\begin{array}{c c}
		0.4 I & 0 \\
		0 & 0.1 I
		\end{array}\right).
\end{equation}

We looked directly at the unitary matrices representing $10$ messages along $E$ as $\lambda_0$ approaches the value $0.4$ from the left. This was actually somewhat of a stroke of luck, since we had earlier convinced ourselves that there is no good reason to look at specific entries appearing in message operators, there having always (previously) been a complete lack of any discernible pattern in these numbers. So it came as quite a surprise that right at the point $\lambda_0=0.4$ where the two boundaries meet, all the unitaries in these sets of $10$ messages were of an easily recognized form (and we generated many such sets, each starting from a different, randomly-generated initial guess). We may always fix one message to be generated by the identity operator (if $U_0\ne I_A$, Alice simply follows all message operators by the unitary $U_0^\dagger$ yielding a new set of messages with $U_0^\prime=I_A$ --- if the original set of messages was valid then this new set will be so, as well), which considered in terms of its four $2$-by-$2$ blocks, is obviously block diagonal. What we have observed is that every set of $10$ unitaries is such that each unitary is paired with one other one in the following sense: first, for every block diagonal unitary, there is another unitary that is block zero-diagonal, by which we mean that both diagonal blocks vanish identically; the remaining unitaries (those that are neither block diagonal or block zero-diagonal) all have the block form,
\begin{eqnarray}\label{Uj}
	U_j=\mu_j\left(\begin{array}{c c}
		u_j & \gamma_j B \\
		-\gamma_j^*B^\dagger u_j & I
		\end{array}\right),
\end{eqnarray}
with
\begin{eqnarray}\label{uj}
	u_j=\left(\begin{array}{c c}
		-x+i\alpha_j & -\beta_j^* \\
		\beta_j & -x-i\alpha_j
		\end{array}\right),
\end{eqnarray}
where $\alpha_j,\beta_j,\mu_j,\gamma_j\in{\cal C}$ and $x\equiv\lambda_2/\lambda_0$. The $u_j$ satisfy
\begin{equation}\label{ujuk}
	\textrm{Tr}(u_ju_k^\dagger)=-2x
\end{equation}
for $j\ne k$, which with $u_0=I$ can be shown to uniquely fix $u_1$ such that $\beta_1=0$. Once the phase of one other $\beta_j$ is chosen, the remaining $u_j$ are also uniquely fixed.

The matrix $B$ appearing in Eq.~(\ref{Uj}) is unitary, independent of $j$, and proportional to the upper-right block of every one of the $U_j$, even those that are block zero-diagonal. In addition, $\gamma_j^*\gamma_k$ is real for every pair $j\ne k$. The pairing of these unitaries that have no zero blocks is such that, with $U_j$ paired with $U_k$, both $u_j=u_k$ and $\gamma_j^*\gamma_k=-1$. For many runs of the numerical search, the $10$ unitaries found are such that only the identity and its partner have zero blocks, the other $8$ being as shown in Eq.~(\ref{Uj}). 

The important thing to note about the pairing described above is this: for those pairs that have the form of Eq.~(\ref{Uj}), it is always possible to take linear combinations to form two new unitaries, one of which is block diagonal,
\begin{eqnarray}\label{blockdiag}
	\left(\begin{array}{c c}
		u_j & 0 \\
		0 & I
		\end{array}\right),
\end{eqnarray}
the other of which is block zero-diagonal,
\begin{eqnarray}\label{bzd}
	\left(\begin{array}{c c}
		0 & B \\
		-B^\dagger u_j & 0
		\end{array}\right).
\end{eqnarray}
In addition, these are obviously $\Lambda$-orthogonal to each other, and by linearity, the entire collection remains so, as well. Therefore, the pair of unitaries, $U_j$ and $U_k$, may be replaced in a set of messages by the above pair of block diagonal and block zero-diagonal unitaries. We therefore believe that,
\begin{allBlock}For $\Lambda=\Lambda_*$, every possible set of $10$ unitary messages is equivalent to a set consisting of $5$ block diagonal and $5$ block zero-diagonal unitaries, all of which are of the forms shown in Eqs.~(\ref{blockdiag}) and (\ref{bzd}). By equivalent we mean in the sense of allowing linear combinations as described above (we also assume that the identity matrix is one of the messages).\end{allBlock}

We note that the set, $\{u_j\}$, appearing in the block zero-diagonal matrices need not be the same as the set appearing in the block diagonal ones, since \textit{any} block diagonal matrix is $\Lambda$-orthogonal to \textit{any} block zero-diagonal one. However, when these sets are not the same, linear combinations of the pairs \textit{cannot} be unitary, which we believe is why the numerical searches always find the two sets to be the same. The reason is that numerical searches rarely find solutions that have zeros in them, unless those are the only solutions available --- the search is much more likely to find the linear combinations, which have few zeros or none at all, rather than those with zero blocks. Actually, this is an argument in favor of conjecture $1$, since the searches \textit{always} find a solution that can, in a very simple way, be transformed into a solution for which at least one-half of the parameters it has found are equal to zero.

If we stay along $E$ but move slowly away from $\Lambda_*$ toward smaller $\lambda_0$, numerically generated solutions continuously depart from the form of Eq~(\ref{Uj}). However, although it is possible to show that $5$ $\Lambda$-orthogonal block diagonal (or block zero-diagonal) such unitaries exist if and only if $\Lambda=\Lambda_*$, it is nonetheless possible to construct $4$ such unitaries everywhere along $E$ as long as $\lambda_0>3/8$. In Appendix~\ref{app:d4}, we prove the following theorem:
\begin{no9U}\label{th:no9U}Suppose we start with $4$ block diagonal and $4$ block zero-diagonal mutually $\Lambda$-orthogonal unitaries of the form given in Eqs.~(\ref{blockdiag}) and (\ref{bzd}), along $E$ with $\lambda_0>3/8$. If it is possible to find a $9^{th}$ $\Lambda$-orthogonal unitary, then it is also possible to find a $10^{th}$, and this is only possible if $\lambda_0\le 0.4$.\end{no9U}

This demonstrates that it is not possible for there to be $9$ unitary messages along $E$ for $\lambda_0>0.4$, at least if $8$ of those unitaries are of the form described in the theorem. Of course, our theorem does not prove that it is impossible to find $9$ unitary messages beyond this point, since we have restricted consideration to these special sets of messages. However, given that the numerical calculations only find solutions that are of this special type, we feel the evidence is quite compelling that in fact $9$ unitary messages do not exist in this region. Assuming this is indeed the case, then since $9$ non-unitary messages definitely do exist for $\lambda_0>0.4$, we may conclude that unitary encoding is not optimal when using the corresponding entangled states for the purpose of dense coding.

\section{Conclusions}
We have investigated the use of unitary and non-unitary operations for deterministic dense coding using both numerical and analytical methods. In dimensions $d=2$ and $d=3$, we have found that non-unitary encoding provides no improvement over what can be accomplished with unitaries. In $d=4$, on the other hand, we have found compelling numerical evidence that this is no longer the case, and we have given analytical arguments that strongly support this view. We therefore conclude that unitary operations are not optimal for deterministic dense coding. Given the wide range of circumstances, discussed in the Introduction, under which unitary encoding allows one to transmit the maximum possible amount of classical information, this leads naturally to the question of whether or not there exist other quantum communication protocols for which unitary encoding is strictly less than optimal.

\begin{acknowledgments}This work has been supported in part by the National Science Foundation through Grant PHY-0456951, and through a grant from the Research Corporation. We are extremely grateful to Shay Mozes for sharing with us the Matlab code he used to generate the numerical results of \cite{Mozes}, and to R.B. Griffiths and his research group for numerous stimulating discussions on this topic.
\end{acknowledgments}

\appendix
\section{Dense coding with two qubits}\label{app:d2}
When Bob and Alice share a non-maximally entangled state of two qubits (dimension $2 \times 2$), it is known that if Alice is restricted to using unitary encoding, they cannot send more than two messages deterministically. Here, we prove that the same conclusion holds even when Alice is allowed to use the most general quantum operations, described by Kraus operators as discussed in the previous section.

\begin{no3}\label{th:no3}Dense coding (by which we mean the ability to send more than $d$ classical messages through a perfect channel assisted by entanglement) is impossible with an entangled resource of dimension $d\times d$ when $d=2$ unless that resource is a maximally entangled state.\end{no3}
Proof: Suppose by contradiction they can succeed with $3$ messages when $\lambda_0 \ne \lambda_1 = 1 - \lambda_0$. The first thing to note is that, by Eq.~(\ref{kappa}) of the main text, at least two of these messages must use unitary encoding by Alice, and the third message, which we already know cannot be unitary, must be encoded by something of Kraus rank equal to $2$. Recall that we may always take the first message to be encoded by the identity operator, $U_0 = I_A$. Hence, the first message is represented by $\vert\Psi_0\rangle$. Let us then write the unitary for the second message as
\begin{equation}
	U_1 = \sum_{n=0}^1 |a_n\rangle_A\langle n|,
\end{equation}
with $\{|a_n\rangle\}$ an orthonormal basis of ${\cal H}_A$. With Eq.~(\ref{Psi0}) defining $\vert\Psi_0\rangle$, we have
\begin{equation}
	|\Psi_1\rangle=(U_1\otimes I_B)|\Psi_0\rangle = \sum_{j=0}^1 \sqrt{\lambda_j} |a_j\rangle_A |j\rangle_B,
\end{equation}
which by orthogonality to $|\Psi_0\rangle$, yields the condition
\begin{equation}
	\label{ortho}
	0 = \sum_{j=0}^1 \lambda_j \langle j|a_j\rangle.
\end{equation}
Writing $|a_0\rangle = \mu |0\rangle + \nu |1\rangle$ and $|a_1\rangle = e^{i\phi}(\nu^* |0\rangle - \mu^* |1\rangle)$, Eq.~(\ref{ortho}) becomes $0 = \mu\lambda_0 - e^{i\phi}\mu^* \lambda_1$, which with $\lambda_0 \ne \lambda_1$, implies $\mu = 0$ and $\nu = e^{i\theta}$ for some real $\theta$. Thus, $U_1 = \nu^*|0\rangle_A\langle 1| + \nu|1\rangle_A\langle 0|$ and 
\begin{equation}
	|\Psi_1\rangle = \nu \sqrt{\lambda_0} |1\rangle_A |0\rangle_B + \nu^* \sqrt{\lambda_1} |0\rangle_A |1\rangle_B.
\end{equation}

Now we need to design a third message, encoded by Kraus operators $K_{2m}$ acting on ${\cal H}_A$, such that each of the states $(K_{2m}\otimes I_B)|\Psi_0\rangle$ is orthogonal to both $|\Psi_0\rangle$ and $|\Psi_1\rangle$ (we here allow an arbitrary number of Kraus operators, though the Kraus rank must of course be equal to $2$). Thus, we may write the states of message $3$ as linear combinations of the following orthonormal basis of the subspace in which they must lie:
\begin{eqnarray}
	|\Phi_1\rangle =  \sqrt{\lambda_1} |0\rangle_A |0\rangle_B -  \sqrt{\lambda_0} |1\rangle_A |1\rangle_B\nonumber \\
	|\Phi_2\rangle = \nu \sqrt{\lambda_1} |1\rangle_A |0\rangle_B - \nu^* \sqrt{\lambda_0} |0\rangle_A |1\rangle_B.
\end{eqnarray}
That is,
\begin{eqnarray}
	(K_{2m}\otimes I_B)|\Psi_0\rangle = \alpha_m|\Phi_1\rangle + \beta_m |\Phi_2\rangle = \sqrt{\lambda_1}(\alpha_m |0\rangle_A + \nu \beta_m |1\rangle)|0\rangle_B - \sqrt{\lambda_0} (\nu^* \beta_m  |0\rangle_A + \alpha_m|1\rangle)|1\rangle_B.
\end{eqnarray}
This determines the operators,
\begin{eqnarray}
	K_{2m} = \sqrt{\frac{\lambda_1}{\lambda_0}}(\alpha_m |0\rangle + \nu \beta_m |1\rangle)_A\langle 0| - \sqrt{\frac{\lambda_0}{\lambda_1}} (\nu^* \beta_m  |0\rangle + \alpha_m|1\rangle)_A\langle 1|,
\end{eqnarray}
so that
\begin{eqnarray}
	K_{2m}^\dagger K_{2m} & = & \frac{\lambda_1}{\lambda_0}(|\alpha_m|^2 + |\beta_m|^2 )|0\rangle_A\langle 0| + \frac{\lambda_0}{\lambda_1}(|\beta_m|^2 + |\alpha_m|^2 )|1\rangle_A\langle 1| \nonumber\\ 
	& - & \nu^*(\alpha_m^*\beta_m + \beta_m^*\alpha_m)|0\rangle_A\langle 1| - \nu(\beta_m^*\alpha_m + \alpha_m^*\beta_m)|1\rangle_A\langle 0|.
\end{eqnarray}
Since $\lambda_0 \ne \lambda_1$, then when we sum this expression over $m$, the resulting diagonal elements will be unequal, implying these operators cannot satisfy the required condition, Eq.~(\ref{complete}), that $\sum_m K_{2m}^\dagger K_{2m} = I_A$. This contradicts our original assumption, and we have therefore proved that with a non-maximally entangled state, $3$ messages is impossible no matter what operations Alice chooses to use.

\section{Proof of theorem \ref{th:no9U}}\label{app:d4}
Here we prove the theorem from Sec.~\ref{sec:d4} of the main text.

{\bf Theorem~\ref{th:no9U}} \textit{Suppose we start with $4$ block diagonal and $4$ block zero-diagonal mutually $\Lambda$-orthogonal unitaries of the form given in Eqs.~(\ref{blockdiag}) and (\ref{bzd}), along $E$ with $\lambda_0>3/8$. If it is possible to find a $9^{th}$ $\Lambda$-orthogonal unitary, then it is also possible to find a $10^{th}$, and this is only possible if $\lambda_0\le 0.4$.}

Proof:
The most general set of $4$ block diagonal $\Lambda$-orthogonal matrices along $E$ (including $U_0=I_A$) is given (with $u_0=I$) by
\begin{equation}
U_j=\left(\begin{array}{c c}
	Au_jA^\dagger & 0\\
	0 & I
\end{array}\right),~~~j=0,1,2,3,
\end{equation}
where $A$ is unitary, and the most general set of block zero-diagonal matrices is
\begin{equation}
U_j=\left(\begin{array}{c c}
	0 & B\\
	B_1v_jB_2 & 0
\end{array}\right),~~~j=4,5,6,7,
\end{equation}
with $B,~B_1$, and $B_2$ all unitary, while the set of matrices $u_j$ have the form given in Eq.~(\ref{uj}) and obey Eq.~(\ref{ujuk}), and similarly for the set $v_j$. One can readily see that the $u_j$ form a linearly independent set, as do the $v_j$.

Let us now consider an arbitrary unitary operator, written in block form as
\begin{equation}
	W=\left(\begin{array}{c c}
		a & b\\
		c & d\end{array}\right).
\end{equation}
Using the linear independence of the two sets, $u_j$ and $v_j$, and requiring $\Lambda$-orthogonality of $W$ with each of the $8$ unitaries, $U_j$, it is straightforward to show that
\begin{eqnarray}
	a=-x\delta\tilde a,\nonumber\\
	c=-x\gamma\tilde a^\prime,
\end{eqnarray}
with $\delta=\textrm{Tr}(d)$ and $\gamma=\textrm{Tr}(bB^\dagger)$,
\begin{equation}
	\tilde a=\frac{1}{2}A\left(\begin{array}{c c}
		1+i\sqrt{\frac{1+x}{1-x}} & e^{i\theta}\sqrt{\frac{1+x}{(1-x)(1-2x)}}\left(1+i\sqrt{\frac{1-x}{1-3x}}~\right)\\
		e^{-i\theta}\sqrt{\frac{1+x}{(1-x)(1-2x)}}\left(1-i\sqrt{\frac{1-x}{1-3x}}~\right) & 1-i\sqrt{\frac{1+x}{1-x}}\end{array}\right)A^\dagger,
\end{equation}
\begin{equation}
	\tilde a^\prime=\frac{1}{2}B_1\left(\begin{array}{c c}
		1+i\sqrt{\frac{1+x}{1-x}} & e^{i\phi}\sqrt{\frac{1+x}{(1-x)(1-2x)}}\left(1+i\sqrt{\frac{1-x}{1-3x}}~\right)\\
		e^{-i\phi}\sqrt{\frac{1+x}{(1-x)(1-2x)}}\left(1-i\sqrt{\frac{1-x}{1-3x}}~\right) & 1-i\sqrt{\frac{1+x}{1-x}}\end{array}\right)B_2,
\end{equation}
and $\theta,~\phi$ are arbitrary real phases. Given these expressions, it is readily seen that
\begin{eqnarray}\label{adaga}
	a^\dagger a=\frac{x^2|\delta|^2}{1-3x}I,\nonumber\\
	c^\dagger c=\frac{x^2|\gamma|^2}{1-3x}I.
\end{eqnarray}

We now impose the constraint that $W$ is unitary. Considering the upper-left blocks of the expression for $I_A=W^\dagger W$, we find that
\begin{equation}\label{delgam}
	|\delta|^2+|\gamma|^2=\frac{1-3x}{x^2}.
\end{equation}
The diagonal blocks of $I_A=WW^\dagger$ yield the restrictions,
\begin{eqnarray}
	I=\frac{x^2|\delta|^2}{1-3x}I+bb^\dagger,\nonumber\\
	I=\frac{x^2|\gamma|^2}{1-3x}I+dd^\dagger.
\end{eqnarray}
These latter equations mean there exist two-by-two unitary matrices $\cal V$ and $\cal V^\prime$ such that
\begin{eqnarray}
	b=\sqrt{1-\frac{x^2|\delta|^2}{1-3x}}{\cal V}B,\nonumber\\
	d=\sqrt{1-\frac{x^2|\gamma|^2}{1-3x}}{\cal V}^\prime.
\end{eqnarray}
Recalling the definitions $\delta=\textrm{Tr}(d)$ and $\gamma=\textrm{Tr}(bB^\dagger)$, taking traces of the preceding equations (after multiplying the first by $B^\dagger$), and then squaring magnitudes gives
\begin{eqnarray}\label{traces}
	|\gamma|^2=\left(1-\frac{x^2|\delta|^2}{1-3x}\right)\left|\textrm{Tr}(\cal V)\right|^2,\nonumber\\
	|\delta|^2=\left(1-\frac{x^2|\gamma|^2}{1-3x}\right)\left|\textrm{Tr}(\cal V^\prime)\right|^2.
\end{eqnarray}
Next, add these two expressions together and use the following facts: (1) the squared magnitude of the trace of a two-by-two unitary cannot exceed $4$; and (2) from Eq.~(\ref{delgam}) the expressions in parentheses on the right-hand side of each of Eqs.~(\ref{traces}) is non-negative. Then, again using Eq.~(\ref{delgam}), we find that
\begin{equation}
	\frac{1-3x}{x^2}\le4,
\end{equation}
from which one directly concludes that $x\ge1/4$ or in other words, $W$ cannot be unitary unless $\lambda_0\le0.4$. That is, given the original set of $4$ block diagonal and $4$ block zero-diagonal matrices of Eqs.~(\ref{blockdiag}) and (\ref{bzd}), it is not possible to find a $9^{th}$ unitary along $E$ unless $\lambda_0\le0.4$.

An alternative method of proof is to show that if $W$ is unitary and $\Lambda$-orthogonal to all the $U_j$, then so is
\begin{equation}
	W^\prime=\left(\begin{array}{c c}
		-x\delta^\prime A\tilde a A^\dagger & b^\prime\\
		-x\gamma^\prime B_1\tilde a^\prime B_2 & d^\prime\end{array}\right),
\end{equation}
and that $b^\prime,~d^\prime$ can be chosen such that $W$ is $\Lambda$-orthogonal to $W^\prime$. Just take $b^\prime=id^\dagger B$ and $d^\prime=-ib^\dagger B$, so that $\delta^\prime=\textrm{Tr}(d^\prime)=-i\gamma^*$ and $\gamma^\prime=\textrm{Tr}(b^\prime B^\dagger)=i\delta^*$. Thus, we may conclude that if it is possible to find a $9^{th}$ unitary it is also possible to find a $10^{th}$, and since $10$ unitary messages is only possible when $\lambda_0\le d/N=0.4$ \cite{ourDense}, we arrive at the same conclusion as before. \hspace{\stretch{1}}$\blacksquare$

%\bibliography{QInfoRefs}
%\bibliographystyle{prsty}

\end{document}